\documentclass{iaus}
\usepackage{epsfig}
\usepackage{floatflt}

  \checkfont{eurm10}
  \iffontfound
    \IfFileExists{upmath.sty}
      {\typeout{^^JFound AMS Euler Roman fonts on the system,
                   using the 'upmath' package.^^J}%
       \usepackage{upmath}}
      {\typeout{^^JFound AMS Euler Roman fonts on the system, but you
                   dont seem to have the}%
       \typeout{'upmath' package installed. iaus.cls can take advantage
                 of these fonts,^^Jif you use 'upmath' package.^^J}%
      }
  \else
  \fi

  \checkfont{msam10}
  \iffontfound
    \IfFileExists{amssymb.sty}
      {\typeout{^^JFound AMS Symbol fonts on the system, using the
                'amssymb' package.^^J}%
       \usepackage{amssymb}%
         \let\leq=\leqslant
         
      }{}
  \fi

  \IfFileExists{amsbsy.sty}
    {\typeout{^^JFound the 'amsbsy' package on the system, using it.^^J}%
     \usepackage{amsbsy}}
    {\providecommand\boldsymbol[1]{\mbox{\boldmath $##1$}}}

\newsavebox{\astrutbox}
\sbox{\astrutbox}{\rule[-5pt]{0pt}{20pt}}

\def\aj{{AJ}}

\def\apj{{ApJ}}
\def\apjl{{ApJL}}
\def\apjs{{ApJS}}

\def\gtrsim{\mathrel{\hbox{\rlap{\hbox{\lower4pt\hbox{$\sim$}}}\hbox{\raise2pt\hbox{$>$}}}}}
\newcommand{\halpha}{H\ensuremath{\alpha}}
\newcommand{\hbeta}{H\ensuremath{\beta}}
\newcommand{\hn}{\halpha+[N{\small II}]}

\newcommand{\kms}{km s\ensuremath{^{-1}}}

\newcommand{\lf}{\ensuremath{L_{5100}}}

\newcommand{\mbh}{\ensuremath{M_\mathrm{BH}}}

\def\mnras{{MNRAS}}

\newcommand{\msigma}{\ensuremath{M_{\mathrm{BH}}-\sigmastar}}
\newcommand{\msun}{\ensuremath{M_{\odot}}}

\newcommand{\sigmastar}{\ensuremath{\sigma_{\star}}}

\title[The Interplay among Black Holes, Stars and ISM in Galactic 
       Nuclei]{Intermediate-mass Black Holes in Galactic Nuclei}

\author[J. E. Greene {\it et al.\/}]%
{Jenny E. Greene$^1$%
Luis C. Ho$^2$\break
\and Aaron J. Barth$^3$}

\affiliation{$^1$Harvard-Smithsonian Center for Astrophysics, 
Cambridge, MA 02138, USA\\[\affilskip]
$^2$The Observatories of the Carnegie Institute of Washington,
Pasadena, CA 91101, USA\\[\affilskip]
$^3$California Institute of Technology, Pasadena, CA, 91125,
USA
}

\pubyear{2004}
\volume{222}
\pagerange{1--8}
\date{?? and in revised form ??}
\jname{The Interplay among Black Holes, Stars and ISM \\in Galactic Nuclei}
\editors{Th. Storchi Bergmann, L.C. Ho \& H.R. Schmitt, eds.}
\begin{document}

\maketitle

\begin{abstract}

We present the first homogeneous sample of intermediate-mass black
hole candidates in active galactic nuclei.  Starting with broad-line
active nuclei from the Sloan Digital Sky Survey, we use the 
linewidth-luminosity-mass scaling relation to select a sample
of 19 galaxies in the mass range $M_{\mathrm{BH}} \approx 8 \times
10^4 - 10^6 \, \msun$.  In contrast to the local active galaxy
population, the host galaxies are $\sim 1$ mag 
fainter than $M^*$ and thus are probably
late-type systems.  The active nuclei are also faint, 
with $M_g \approx -15$ to $-18$ mag, while the bolometric luminosities
are close to the Eddington limit.  The spectral properties of the 
sample are compared to the related class of objects known as
narrow-line Seyfert 1 galaxies.  We discuss the importance of our
sample as observational analogues of 
primordial black holes, contributors to the 
integrated signal for future gravitational wave experiments, and as a
valuable tool in the calibration of the \msigma\ relation.

\end{abstract}

\firstsection 
\section{Background \& Sample Selection}

Dynamical studies have established the existence of supermassive black
holes (BHs) with masses $M_{\rm BH} \approx 10^6-10^9$~\msun\ in
the centers of most, if not all, local galaxies with bulges (Magorrian
et al.  1998; Richstone 2004).  A significant challenge to any model
of cosmological BH growth is the nature of seed BHs.  Observations of
intermediate-mass BHs ($M_{\rm BH} \approx 10^3-10^6$~\msun) in
the local Universe would provide the most direct empirical constraints
on such seeds.  Their number and mass
distribution will impact the expected integrated background in
gravitational radiation (e.g., Hughes 2002).  
Intermediate-mass BHs have practical value in that they
offer tremendous leverage in anchoring local BH-galaxy scaling 
correlations such as the \msigma\ relation.

Recently discovered intermediate-mass BH candidates in galactic nuclei
offer tantalizing hints as to the nature of this population.  NGC 4395
is a very late-type (Sdm) spiral with no bulge component.
Nevertheless, it has the emission properties
of a type 1 active galactic nucleus (AGN; Filippenko \& Sargent 1989;
Filippenko, Ho, \& Sargent 1993).
Mass estimates based on the \hbeta\ linewidth-luminosity-mass
scaling relation and X-ray variability suggest a BH mass of $10^4 -
10^5$~\msun.  Interestingly, these agree with
the limit of $< 10^5$~\msun\ derived from the \msigma\ relation
(Filippenko \& Ho 2003; Tremaine et~al. 2002).  
POX 52, whose optical spectrum is remarkably
similar to that of NGC 4395, has a dwarf
elliptical host (Barth et al. 2004).  The galaxy has a
central velocity dispersion of 36~\kms, which yields a mass estimate
of $M_{\rm BH} = 1.4 \times 10^5$~\msun, again consistent with the
value of $1.6 \times 10^5$ \msun\ derived from the \hbeta\
linewidth-luminosity-mass scaling relation.  NGC 4395 and POX 52
fall on the \msigma\ relation while
their hosts are anomalous compared to the local AGN population
(Ho, Filippenko, \& Sargent, 1997a, 2003; Kauffmann et al. 2003), 
which is invariably
affiliated with massive, bulge-containing galaxies.

Unfortunately, it is currently technically impossible to obtain direct
mass measurements for $M_{\rm BH} \lesssim 10^6$ \msun, because of our
inability to resolve the BH sphere of influence for all but the
nearest galaxies in the Local Group (e.g., M33; Gebhardt et al. 2001).
Thus, the best hope to find intermediate-mass BHs is through an AGN
survey.  However, it may be that objects like NGC 4395 are quite rare.  
To accumulate a significant sample will require a large survey.
Also, both the radiative signatures of accretion and the host
galaxies themselves will be intrinsically faint. As a result,
detection of the faint AGN will be limited by sensitivity and host
galaxy contamination.


\begin{figure}
\begin{center}
\epsfig{file=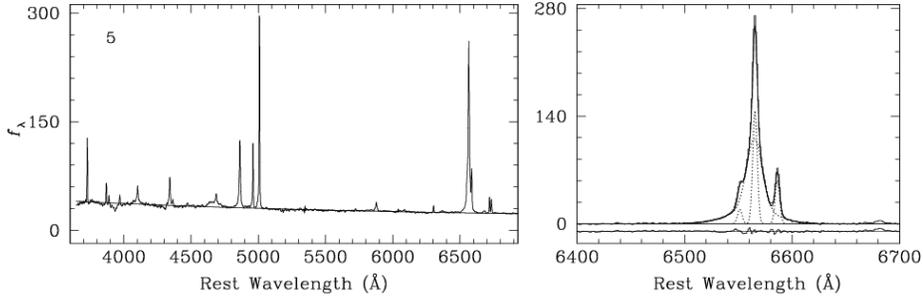,angle=-90,width=0.9\textwidth,keepaspectratio=true}
\vskip -0.1mm
\caption{ 
Sample spectrum.  {\it Left:} An example of our power-law fit
to the continuum.  {\it Right:} Enlargement of the region around
\halpha+[N~{\tiny II}] $\lambda\lambda$6548, 6583.  In the upper plot,
the data are shown in histogram, the individual models for the narrow
and broad lines are shown in dotted lines, and the sum of all the
fitted components is shown as a solid line.  The lower plot gives the
residuals of the fit.  The ordinate of the plots are in units of
$10^{-17}$ erg s$^{-1}$ cm$^{-2}$ \AA$^{-1}$. 
\label{obj5}}
\vskip -3.mm
\end{center}
\end{figure}


A large-area, sensitive, and uniform optical spectroscopic galaxy
survey, such as the Sloan Digital Sky Survey (SDSS; York et~al. 2000), 
offers the best
opportunity for finding a significant number of new intermediate-mass
BH candidates. We present our initial sample from the SDSS First Data
Release (DR1; Abazajian et~al. 2003).

\begin{figure*}
\begin{center}
\epsfig{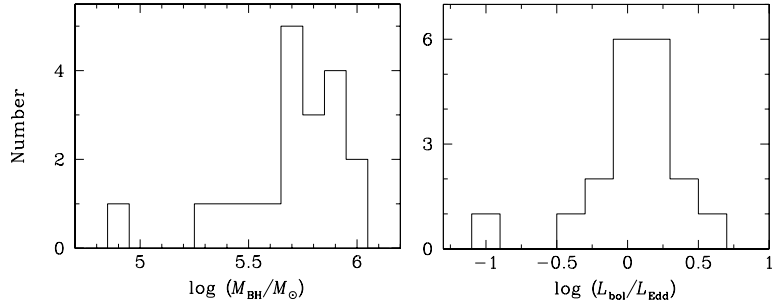}
\vskip -0.1mm
\caption{ 
Distribution of virial BH masses (left) and Eddington ratios
(right) for our sample.
\label{bhmass}}
\vskip -3mm
\end{center}
\end{figure*}

We begin with the 153,000 galaxy and quasar spectra from the DR1.  We
use the Principal Component Analysis code of Hao (2004) to model and remove
contaminating starlight. We then select all broad-line AGN with $z \leq
0.35$ based on the presence of broad \halpha\ emission (Ho
et~al. 1997b).  We estimate BH ``virial'' masses using a variant of
the empirically determined linewidth-luminosity-mass scaling relation
of Kaspi et al. (2000), which relates $M_{\rm BH} \propto L^{0.7}
\upsilon^2$, where $L$ is the AGN luminosity at 5100 \AA\ and
$\upsilon$ is the FWHM of (typically) broad \hbeta.  
To determine $v$ we use the stronger \halpha\ as a surrogate.  To properly
deconvolve the narrow components of the \hn\ complex we model the
[S~{\small II}] $\lambda \lambda$6716, 6731 doublet with a
superposition of Gaussian components and then scale this model to fit
the complex (Ho et al. 1997b).  The continuum luminosity is derived
from a power-law fit (see Fig. \ref{obj5}).  From the sample of
objects with \mbh$\leq 10^6$~\msun\ we remove all objects with a
stellar continuum $\gtrsim 20\%$ of the total continuum, due to
unreliable \halpha\ FWHM measurements. The final sample of 19 objects
has a mass distribution shown in Fig. \ref{bhmass} (for details see
Greene \& Ho 2004).


\section{Properties}

Our objects are analogous to the class of AGNs known as narrow-line
Seyfert 1s (NLS1), which are thought to possess relatively low-mass
BHs emitting at a high fraction of their Eddington rates (e.g.,
Boroson 2002).  By the
linewidth criterion, all of the objects would technically
qualify as ``narrow-line'' sources: the broad \halpha\ component has
FWHM ranging from 464 to 1730 \kms. In keeping with expectation the derived
Eddington ratios mostly cluster around $L_{\rm bol}/L_{\rm Edd}
\approx 1$ (Fig. \ref{bhmass}).  
We caution that these values
were derived assuming a single bolometric correction of 
$L_{\rm bol}$ = 9.8\lf\ (McLure \& Dunlop 2004) although the spectral
energy distributions of AGNs are known to vary significantly with
accretion rate (Pounds et al. 1995; Ho 1999).
However, our sample shows a greater diversity of
Fe~{\small II} and [O~{\small III}] strengths than in previous samples of
NLS1s. 

\begin{floatingfigure}[r]{1.7in}
\begin{center}
\epsfig{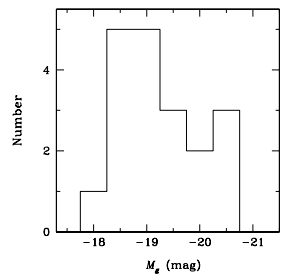}
\vskip -1.mm
\caption{
Host galaxy $g$-band absolute magnitudes.  
\label{maghist}}
\vskip -3.mm
\end{center}
\end{floatingfigure}

The morphology of the host galaxies of intermediate-mass BHs is of
fundamental importance for understanding the origin of this class of
objects and their relationship to the overall demography of central
BHs in galaxies. While we cannot obtain detailed morphology from the
SDSS data, the host absolute magnitudes (Fig.
\ref{maghist}) are relatively low-luminosity and therefore more likely
to be late-type galaxies (c.f. NGC 4395 and POX 52 above).  
More detailed morphological information requires deep, 
high-resolution imaging to disentangle a (presumably tiny) bulge component.

\section{Future Work}

\begin{figure}
\begin{center}
\epsfig{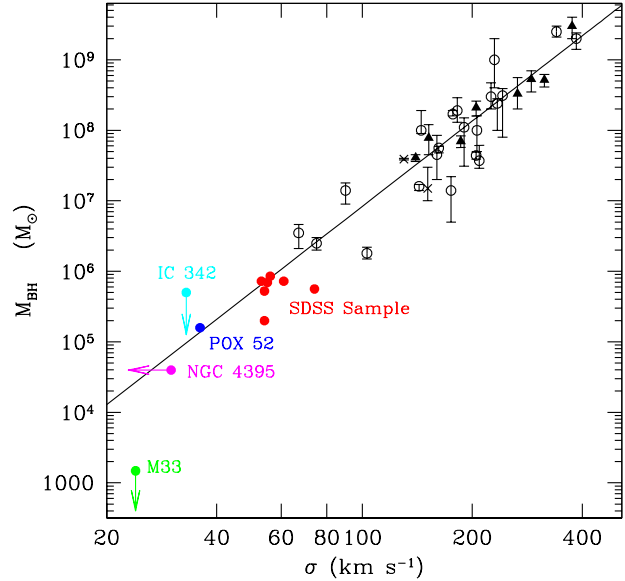}
\vskip -0.1mm
\caption{
The \msigma\ relation, including objects from our February ESI
run. 
Points with error bars represent nearby galaxies with fundamental
measurements of black hole mass from stellar dynamics (open circles),
gas dynamics (triangles), or maser dynamics (crosses).  Upper limits
to the black hole masses in the nearby, inactive galaxies M33
(Gebhardt et~al. 2001) and IC 342 (B{\" o}ker et~al 1999) as well as
NGC 4395 and POX 52 are shown. This preliminary result suggests our
objects are consistent with an extrapolation of the local \msigma\ relation. 
\label{msigma}}
\vskip -2mm
\end{center}
\end{figure}

We employ a simple technique to exploit the breadth of the SDSS galaxy
spectroscopy to extend the demography of central BHs at least 1 order
of magnitude below the $10^6$ \msun\ threshold, a regime hardly
explored previously.  We hope to use the width of narrow emission
lines as a proxy for stellar velocity dispersion (e.g., Nelson \&
Whittle 1996) to select additional candidates .  Preliminary stellar
velocity dispersions using the ESI spectrograph at Keck yielded 12 AGNs
with \sigmastar\ $<$ 70 km s$^{-1}$.  As Fig. \ref{msigma} shows,
the objects with velocity dispersions appear to fall on the
extrapolation of the local \msigma\ relation.
The complete data set will test
whether the BH mass estimates we derive obey the \msigma\ relation
and the reliability of the virial mass estimator for AGNs.  It
would also be valuable to compare the velocity dispersions derived
from stars with those derived from gas.  High-resolution, deep imaging
is needed to quantify the morphologies and detailed structural
parameters of the host galaxies.  Do these objects have bulges, and if
so, what kind and where do they fall on galaxy scaling relations like
the fundamental plane?  Finally, much work remains to be done,
especially at non-optical wavelengths, to further characterize the
properties of the AGNs themselves.


\begin{acknowledgments}
J.~E.~G. acknowledges a AAS international
travel grant. We are grateful to Lei Hao for making available
her PCA software, and to the entire SDSS
collaboration for providing the extraordinary database and processing
tools that made this work possible.

\end{acknowledgments}


\begin{thebibliography}{}

\bibitem[]{}Abazajian, K., et~al. 2003, \aj, 126, 2081

\bibitem[]{}Barth, A.~J., Ho, L.~C., Rutledge, R.~E., \& 
Sargent, W.~L.~W. 2004, \apj, in press (astro-ph/0402110)

\bibitem[]{} B{\" o}ker, T., van der Marel, R.~P., \& Vacca, 
W.~D.\ 1999, \aj, 118, 831 

\bibitem[]{}Boroson, T.~A. 2002, \apj, 565, 78

\bibitem[]{}Filippenko, A.~V. \& Ho, L.~C. 2003, \apjl, 588, L13

\bibitem[]{}Filippenko, A.~V., Ho, L.~C., \& Sargent, W.~L.~W. 1993, \apjl, 410, L75

\bibitem[]{}Filippenko, A.~V. \& Sargent, W.~L.~W. 1989, \apjl, 342, L11

\bibitem[]{}Gebhardt, K., et~al.  2001, \aj, 122, 2469

\bibitem[]{}Greene, J.~E., \& Ho, L.~C. 2004, \apj, in press 
(astro-ph/0404110)

\bibitem[]{}Hao, L. \& Strauss, M. 2004, in Carnegie Observatories
  Astrophysics Series, Vol. 1: Coevolution of Black Holes and
  Galaxies, ed. L. C. Ho (Pasadena:Carnegie Observatories), in press

\bibitem[]{}Ho, L.~C. 1999, \apj, 516, 672

\bibitem[]{}Ho, L.~C., Filippenko, A.~V., \& Sargent, W.~L.~W. 1997a, \apj, 487, 568

\bibitem[]{}------. 2003, \apj, 583, 159

\bibitem[]{}Ho, L.~C., Filippenko, A.~V., Sargent, W.~L.~W., \& Peng,
C.~Y. 1997b, \apjs, 112, 391

\bibitem[]{}Hughes, S.~A. 2002, \mnras, 331, 805

\bibitem[]{}Kaspi, S., Smith, P.~S., Netzer, H., Maoz, D., Jannuzi, B.~T., \&
  {Giveon}, U. 2000, \apj, 533, 631

\bibitem[]{}Kauffmann, G., et al. 2003, \mnras, 346, 1055


\bibitem[]{}Magorrian, J. et~al. 1998, \aj, 115, 2285

\bibitem[]{}McLure, R.~J., \& Dunlop, J.~S. 2004, \mnras, in press
  (astro-ph/0310267)

\bibitem[]{}Nelson, C.~H., \& Whittle, M. 1996, \apj, 465, 96

\bibitem[]{}Pounds, K.~A., Done, C., \& Osborne, J.~P. 1995, \mnras, 277, L5

\bibitem[]{}Richstone, D. 2004, in Carnegie Observatories Astrophysics
Series, Vol. 1: Coevolution of Black Holes and Galaxies, ed. L. C. Ho
(Cambridge: Cambridge Univ. Press), in press

\bibitem[]{}Tremaine, S., et~al. 2002, \apj, 574, 740

\bibitem[]{}York, D.~G., et~al., 2000, \aj, 120, 1579

\end{thebibliography}
\end{document}